\documentclass[12pt]{article}

\usepackage[tmargin=2cm,lmargin=2.5cm,bmargin=3cm,hmarginratio=1:1,headheight=65pt,headsep=1cm]{geometry} 
\usepackage[T1]{fontenc}
\usepackage[french,english]{babel}
\usepackage{color,graphicx}
\usepackage{amsmath,amssymb,amsfonts}
\usepackage{amsthm}
\usepackage{mathtools}
\usepackage{slashed}
\usepackage{sidecap} 
\usepackage{array}
\usepackage[indentafter]{titlesec}
\usepackage[utf8]{inputenc}
\usepackage{setspace}
\usepackage{perpage} 
\usepackage{fancyhdr} 
\usepackage{emptypage}
\usepackage{pifont} 
\usepackage{cite} 

\usepackage{hyperref} 

\usepackage{color}

\newcommand{\badat}{\begin{alignedat}}
\newcommand{\eadat}{\end{alignedat}}

\newcommand{\p}{\partial}
\newcommand{\rd}{\mathcal}

\newcommand{\ep}{\epsilon}

\def\scri{{\mathcal{I}}}
\def\hor{{\mathcal{H}}}

\def\cH{{\cal H}}
\def\cI{{\cal I}}

\def\cL{{\cal L}}

\def\cO{{\cal O}}

\def\p{\partial}

\date{}
\begin{document}

\begin{titlepage}
  \thispagestyle{empty}

  \begin{flushright}
  CPHT-RR021.042018
  \end{flushright}

\vskip1cm

  \begin{center}  
{\LARGE\textbf{Black hole memory effect}}

\vskip1cm

   \centerline{Laura Donnay$^{a,b}$, Gaston Giribet$^{c,d}$, Hern\'an A. Gonz\'alez$^{e}$, Andrea Puhm$^{a,b,f}$}

\vskip1cm

{$^a$ Center for the Fundamental Laws of Nature, Harvard University}\\
{{\it 17 Oxford Street, Cambridge, MA 02138, USA.}}

{$^b$ Black Hole Initiative, Harvard University}\\
{{\it 20 Garden Street, Cambridge, MA 02138, USA.}}

{$^c$ Physics Department, University of Buenos Aires and IFIBA-CONICET}\\
{{\it Ciudad Universitaria, Pabell\'on 1, Buenos Aires, 1428, Argentina.}}

{$^d$ Center for Cosmology and Particle Physics, New York University,}\\
{{\it 726 Broadway, New York, NY 10003, USA.}}

{$^e$ Institute for Theoretical Physics, TU Wien, }\\
{{\it Wiedner Hauptstr. 8-10, 1040 Vienna, Austria.}}

{$^f$ CPHT, Ecole Polytechnique, CNRS, }\\
{{\it 91128, Palaiseau, France.}}

\end{center}

\vskip1cm

\begin{abstract}
We compute the memory effect produced at the black hole horizon by a transient gravitational shockwave. As shown by Hawking, Perry, and Strominger (HPS) such a gravitational wave produces a deformation of the black hole geometry which from future null infinity is seen as a Bondi-Metzner-Sachs (BMS) supertranslation. This results in a diffeomorphic but physically distinct geometry which differs from the original black hole by their charges at infinity. Here we give the complementary description of this physical process in the near-horizon region as seen by an observer hovering just outside the event horizon. From this perspective, in addition to a supertranslation the shockwave also induces a horizon superrotation. We compute the associated superrotation charge and show that its form agrees with the one obtained by HPS at infinity. In addition, there is a supertranslation contribution to the horizon charge, which measures the entropy change in the process. We then turn to electrically and magnetically charged black holes and generalize the near-horizon asymptotic symmetry analysis to Einstein-Maxwell theory. This reveals an additional infinite-dimensional current algebra that acts non-trivially on the horizon superrotations. Finally, we generalize the black hole memory effect to Reissner-Nordstr\"{o}m black holes.

\end{abstract}

\end{titlepage}

\newpage

\section{Introduction}

Over the last few years we have learned that gravitational and gauge field dynamics in asymptotically Minkowski spacetime entails a rich mathematical structure whose relevance for physics had been largely overlooked. This observation led to a revision of the notion of vacua in gravity and gauge theories in asymptotically flat spacetimes, which is of crucial importance for the scattering problem. This mathematical structure, expressed in the emergence of an infinite set of symmetries, unveils a surprising connection among three previously known but seemingly disconnected topics: {a}) the soft theorems for the S-matrix of gravity and gauge theories in asymptotically flat spacetimes \cite{Weinberg:1965nx}, {b}) the enhanced symmetry group that governs the dynamics in the asymptotic region \cite{Bondi:1962px, Sachs:1962wk, Sachs:1962zza}, and {c}) the memory effect produced by transient gravitational waves \cite{osti_4274059,Christodoulou:1991cr}.

Recently, there has been substantial progress in understanding the three components of this triangle and how they are interconnected. The new insights raise the hope for a better understanding of scattering processes, especially when gravitons, or even black holes are involved. Besides, they could lead to a better comprehension of the physical meaning of the infinite-dimensional symmetries exhibited by Minkowski spacetime in its asymptotic domain. 

The existence of a set of infinite-dimensional asymptotic symmetries in the future (and past) null infinity region(s) has been known for a long time \cite{Bondi:1962px, Sachs:1962wk, Sachs:1962zza}. However, only recently the importance of these symmetries has been understood and significant advances have been made \cite{Barnich:2009se, Barnich:2011mi,Barnich:2010eb,Barnich:2013axa, Strominger:2013jfa, He:2014laa, Cachazo:2014fwa, Kapec:2014opa, Lysov:2014csa, Strominger:2014pwa, Pasterski:2015tva, Bieri:2013hqa, Pasterski:2015zua, Pate:2017fgt}. The algebra generated by these symmetries, known as Bondi-Metzner-Sachs (BMS) algebra, originally appeared in the study of classical gravitational radiation in asymptotically flat spacetimes. The application of the BMS symmetry algebra to study the S-matrix in flat spacetime has been proposed recently \cite{Strominger:2013jfa} and, since then, several physical systems have been studied within this framework; see \cite{Strominger:2017zoo} and references therein and thereof. In particular, a revision of the problem of formation and evaporation of black holes has recently been initiated in \cite{Hawking:2015qqa, Hawking:2016msc, Strominger:2017aeh}, where it was suggested that the BMS symmetry would be of importance for the information loss problem. 

Here, we will not address the information loss puzzle, but rather another problem which is related to the black hole memory: To understand how the BMS symmetry that underlies scattering processes involving a black hole can be measured by an observer hovering just outside the event horizon. We will establish a connection between the description of the black hole geometry in terms of the BMS symmetries in the asymptotic region at null infinity and its description in terms of the symmetries that emerge in the near-horizon region by computing the gravitational memory effect in the vicinity of the black hole horizon produced by an incoming shockwave. We will refer to it as the black hole memory effect \cite{Hawking:2016sgy}. A similar process has recently been studied by Hawking, Perry, and Strominger (HPS) \cite{Hawking:2016sgy}, who showed that a transient shockwave produces a disturbance in the spacetime corresponding to a BMS supertranslation at null infinity. This provides a concrete example of a physical process that endows a black hole with BMS hair of the type suggested in \cite{Hawking:2016msc}. However, it remained an open question how this phenomenon is seen from the point of view of an observer close to the horizon. Here, we will show that the BMS supertranslation hair of \cite{Hawking:2016sgy} can be understood as a supertranslation\footnote{Supertranslations on the future horizon of Schwarzschild black hole have been also studied in \cite{Hawking:2016msc}, where the canonical construction of the Bondi-gauge preserving supertranslations was given.} composed with a superrotation from the point of view of the near-horizon geometry. We compute the conserved charge associated to this superrotation and find its form to agree with the one of HPS. 

This is relevant for several reasons: First, it describes a physical process whose effect on the horizon geometry can be captured by the symmetries discovered in \cite{Donnay:2015abr}. Second, this shows that, in addition to horizon supertranslations \cite{Hawking:2015qqa}, horizon superrotations are crucial to describe the physics in the vicinity of the black hole. Third, this gives a bulk complementary description of the process studied in \cite{Hawking:2016sgy}, which sheds light on the connection between the symmetries emerging in different regions of the spacetime.

We begin in Section 2 with a review of the results of \cite{Hawking:2016sgy} of how the action of the BMS symmetries on the Schwarzschild geometry can be understood as a perturbation produced by a transient gravitational wave. In Section 3, we describe the same physical process from the point of view of an observer in the near-horizon region using the symmetry analysis of \cite{Donnay:2015abr, Donnay:2016ejv}. We compute the superrotation and supertranslation charges associated to the asymptotic symmetries, and we show how the charges found in the near-horizon region relate to those computed at null infinity. In Section 4, we extend the analysis to the case of electrically and magnetically charged black holes. As a prerequisite, we first generalize the near-horizon asymptotic symmetry analysis of \cite{Donnay:2015abr, Donnay:2016ejv} to Einstein-Maxwell theory. This is shown to yield an additional infinite-dimensional current algebra on which the horizon charges of the gravity sector act non-trivially. We discuss the physical interpretation of the extended symmetry and of their associated charges by analyzing the particular case of non-extremal Reissner-Nordstr\"{o}m black holes. We evaluate the zero-modes of the charges on the dyonic solution and discuss their interpretation. The extremal case, which exhibits qualitatively different features, is analyzed separately. In Section 5, we generalize the black hole memory effect to the Reissner-Nordstr\"{o}m black hole. Section 6 contains our conclusions.

\section{Gravitational shockwaves and BMS hair}\label{sec:HPS}

We begin by reviewing the salient features of the BMS analysis of \cite{Hawking:2016sgy} and their proposal for a dynamical mechanism for generating BMS hair on black holes. 

Consider a static Schwarzschild black hole whose line element in advanced Bondi coordinates $(v,r,z^A)$ is given by
\begin{equation}\label{baldSchw}
d{s}^2_0={g}^0_{\mu \nu }dx^{\mu}dx^{\nu}= -\Big(1-\frac{2M}{r}\Big)dv^2+2dvdr+r^2\gamma_{AB}dz^Adz^B, 
\end{equation}
where $v$ is the advanced time and $z^A$ ($A=1,2$) represents the angular position on the 2-sphere with unit metric $\gamma_{AB}$. The horizon is located at $r_+=2M$. At past null infinity $\scri^-$, which is defined as the null surface obtained by taking the limit $r\to \infty$ while keeping $(v,z^A)$ fixed, the only non-vanishing conserved charge is the mass $M$ and hence~\eqref{baldSchw} represents a \emph{bald} Schwarzschild black hole.
In~\cite{Hawking:2016sgy}, HPS constructed BMS supertranslation hair at null infinity and showed that a physical process for the bald Schwarzschild black hole to acquire such hair is given by perturbing the geometry~\eqref{baldSchw} with a linearized gravitational shockwave prepared at advanced time $v_0$ whose energy density to leading order in large radial distance $r$ is given by
\begin{equation}\label{TvvLO}
T_{vv}=\frac{\mu + T(z)}{4\pi r^2}\delta (v-v_0)\,.
\end{equation}
The function $T(z)$ characterizes the angular profile of the shockwave and, following~\cite{Hawking:2016sgy} we explicitly write its monopole contribution $\mu$ separately. 
Solving the conservation equation for the stress-tensor in the background~\eqref{baldSchw} yields the subleading contributions to~\eqref{TvvLO} which break spherical symmetry, namely
\begin{equation}\label{TasymNLO}
T_{vv}= \Big( \frac{\mu + T(z)}{4\pi r^2} +\frac{D_AT^A(z)}{4\pi r^3}\Big) \delta (v-v_0) \,, \quad  T_{vA}= \frac{T^A(z)}{4\pi r^2} \delta (v-v_0) \,.
\end{equation}
Here, $T^A$ obeys the equation $(D^2+2)D_AT^A = -6M T$, with $D_A$ and $D^2\equiv \gamma^{AB}D_A D_B$ being the covariant derivative and the Laplacian on the 2-sphere, respectively.
As explained in~\cite{Hawking:2016sgy}, the solution to these equations can be conveniently expressed in terms of the Green function $G(z,w)$ connecting two different angular positions $z^A$ and $w^{A}$ as defined by
\begin{equation}
D^2(D^2+2)G(z,w)=\frac{4}{\sqrt{\gamma}}\ \delta^{(2)}(z-w)\,,
\end{equation}
where $\gamma$ denotes the determinant of the metric $\gamma_{AB}$ on the sphere. Defining 
\begin{equation}
 C(z)=\int d^2w \ G(z,w)\ T(w) \,,
\end{equation}
the stress-tensor components~\eqref{TasymNLO} become
\begin{equation}\label{TasymCNLO}
\badat{2}
&T_{vv}= \frac{1}{4\pi r^2}\Big( \mu +\frac 14 D^2(D^2+2)C - \frac{3M}{2r}D^2C\Big)\delta(v-v_0) ,\\
&T_{vA} = -\frac{3M}{8\pi r^2}D_AC\ \delta(v-v_0) .
\eadat
\end{equation}
This represents the energy-momentum contribution of the linearized shockwave. Its effect on the background is to produce a perturbed metric $g_{\mu \nu }={g}^0_{\mu \nu }+h_{\mu \nu }$ with the perturbation $h_{\mu \nu}$ given by
\begin{eqnarray}\label{hsupertranslation}
\badat{3}
&h_{vv}= \Big( \frac{2\mu }{r} - \frac{M}{r^2}D^2C\Big)\ \Theta(v-v_0) , \\
&h_{vA}= D_A\Big( \frac{r-2M}{r}C + \frac{1}{2}D^2C\Big) \ \Theta(v-v_0), \\
&h_{AB}= -2r\Big( D_AD_B C- \frac{1}{2}\gamma_{AB} D^2C\Big) \ \Theta(v-v_0). 
\eadat
\end{eqnarray} 
This perturbation was shown in~\cite{Hawking:2016sgy} to be equivalent to acting on the Schwarzschild geometry~\eqref{baldSchw} with a large diffeomorphism $h_{\mu\nu}=\mathcal{L}_{\zeta }{g}^0_{\mu \nu}$ generated by the asymptotic BMS Killing vector 
\begin{equation}
 \zeta=\zeta^{v}\partial_v+\zeta^{A}\partial_A+\zeta^{r}\partial_r \,,
\end{equation}
with components
\begin{equation}\label{zetas}
\zeta^v = f\,, \quad \zeta^{A}=\frac{1}{r}D^Af\partial_A\,, \quad \zeta^{r}=-\frac{1}{2}D^2f\,,
\end{equation}
and where $f=-C(z)\Theta(v-v_0)$; see~\cite{Hawking:2016sgy} for more details. This large diffeomorphism corresponds to a supertranslation that changes the BMS (superrotation) charges at null infinity. 
The resulting supertranslated black hole metric takes the form\footnote{The extension of the supertranslated geometry into the bulk is gauge dependent. Here it is done by requiring the Bondi gauge to be preserved~\cite{Hawking:2016sgy}.}
\begin{equation}
\badat{2}\label{supertranslatedSchw}
ds^2&={g}_{\mu \nu }dx^{\mu}dx^{\nu}=\Big(\frac{2M}{r}-1+\frac{M}{r^2}D^2 f\Big) dv^2+2dvdr-D_A\Big(2f-\frac{4M}{r}f+D^2f\Big)dvdz^A \\
&\qquad \qquad \qquad+\Big(r^2\gamma_{AB}+2rD_AD_Bf-r\gamma_{AB}D^2f\Big) dz^Adz^B \,.
\eadat
\end{equation}
The location of the supertranslated event horizon is 
\begin{equation}
 (r_+)_f=r_{+} +\frac{1}{2}D^2f\,,
\end{equation}
and thus depends on the angular variables through $D^2f$. Note that the solution~\eqref{supertranslatedSchw} is exact in $r$ but only linear\footnote{See~\cite{Barnich:2016lyg} for finite BMS transformations at null infinity.} in $f$ and therefore has to be understood up to order $\mathcal{O}(f^2)$.
The supertranslated Schwarzschild black hole~\eqref{supertranslatedSchw} is a different physical configuration than the unperturbed bald geometry~\eqref{baldSchw} as it carries non-vanishing superrotation charge~\cite{Hawking:2016msc, Hawking:2016sgy} 
\begin{equation}
\badat{2}
Q_Y^{\text{HPS}} = \frac{1}{8\pi }\int_{{\cI^{-}_{+}}} d^2z \sqrt{\gamma}\ Y^A{N_A} =-\frac{3M}{8\pi }\int_{{\cI^{-}_{+}}}d^2z\sqrt{\gamma}\ Y^A\partial_Af , \label{QHPS}
\eadat
\end{equation}
where $Y^A$ is any smooth vector field on the sphere, $N_A$ is the angular momentum aspect, and ${\mathcal{I}^{-}_{+}}$ is the 2-sphere that represents the remote future of past null infinity\footnote{The antipodal matching condition~\cite{Kapec:2014opa} relates the field configurations at $\scri^-_+$ to those at $\scri^+_-$, the latter corresponding to the 2-sphere in the remote past of future null infinity $\scri^+$. We refer to~\cite{Hawking:2016sgy} for the details about the prescription for the matching conditions and integration.} $\mathcal{I}^{-}$.

\begin{figure}[ht!]
\begin{center}
\includegraphics[width=5in]{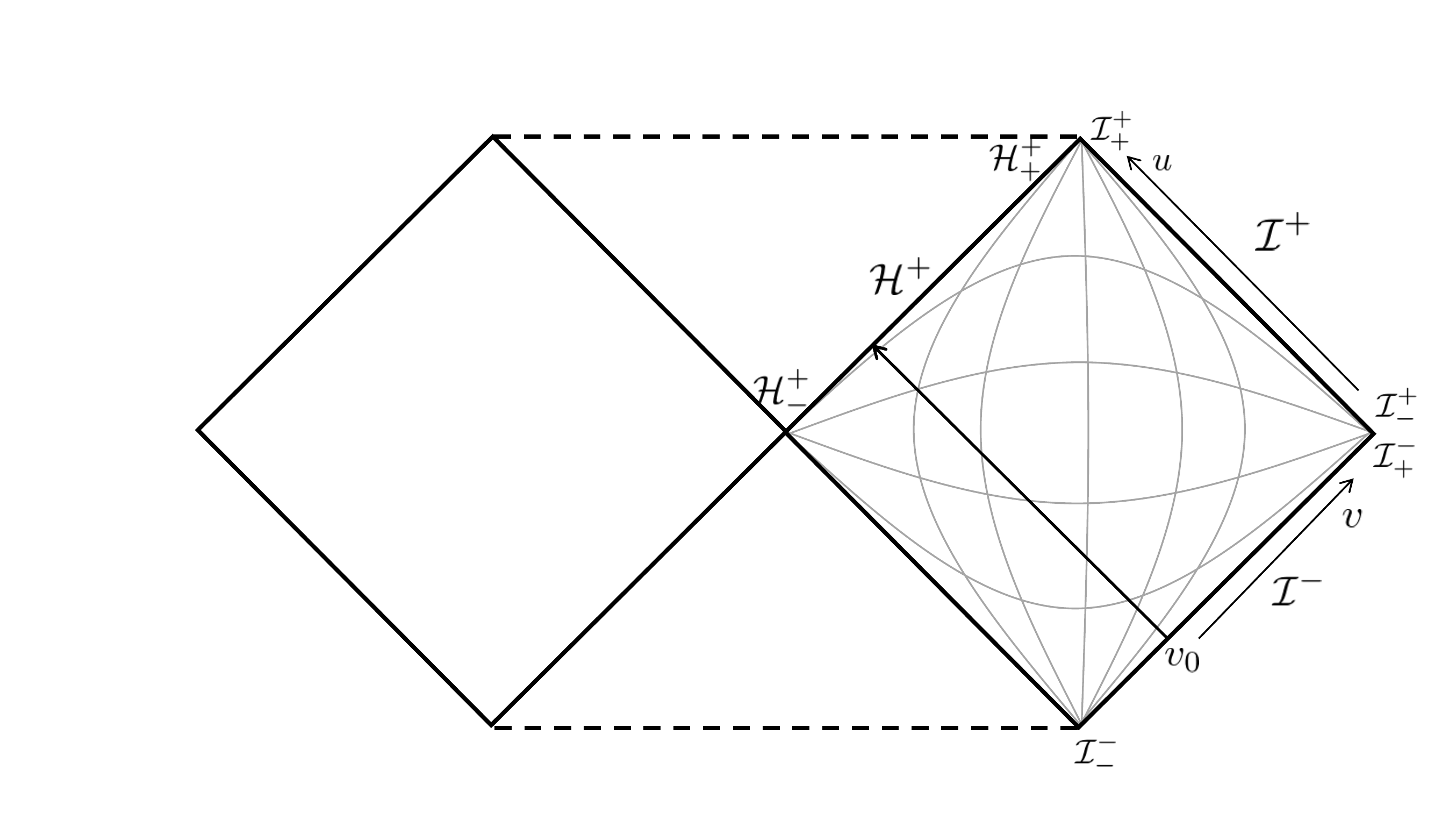}
\caption{Penrose diagram of a Schwarzschild black hole. The gravitational shockwave at $v=v_0$ describes a domain wall that divides the exterior geometry in two regions, each of them with different values of the asymptotic charges.}
\label{Figaro}
\end{center}
\end{figure}

For $v<v_0$ the spacetime is described by the bald Schwarzschild black hole~\eqref{baldSchw}. The perturbation by the shockwave at $v=v_0$ turns on non-vanishing superrotation charge~\eqref{QHPS} and for $v>v_0$ the spacetime is described by the supertranslated Schwarzschild geometry~\eqref{supertranslatedSchw}. See figure~\ref{Figaro}. 
The action of large diffeomorphisms corresponding to supertranslations on the Schwarzschild geometry at null infinity can thus be understood as the physical process of sending in a gravitational shockwave with an asymmetric angular profile. This concludes our review of~\cite{Hawking:2016sgy} and their interpretation of the action of BMS transformations at null infinity. We now turn to the horizon.

\section{Soft hair on Schwarzschild horizons}\label{sec:MemorySchw}

We now discuss how the process of deforming the black hole geometry by an incoming shockwave is seen by an observer located close to the horizon and, moreover, how supertranslations at null infinity get encoded in the symmetry transformations at the horizon. Since the supertranslated black hole solution~\eqref{supertranslatedSchw} is valid for finite values of the radial distance $r$, we can investigate this question using the near-horizon analysis of~\cite{Donnay:2015abr,Donnay:2016ejv}.
There, it was shown that if one starts from the general form of a near-horizon metric
\begin{equation}\label{NHmetric}
ds^2 = -2 \rho \kappa \,dv^2+2 \,dvd\rho +2\rho \theta_A\, dvdz^A+\left(\Omega_{AB}+\rho\lambda_{AB}\right) \,dz^Adz^B+\dots\,,
\end{equation}
where the horizon is located at $\rho=0$, the ellipsis stand for $\cO(\rho^2)$ terms, $\kappa\,,\theta_A\,,\Omega_{AB}\,, \lambda_{AB}$ are in principle arbitrary functions\footnote{The function $\lambda_{AB}$ does not ultimately appear in the conserved charges~\cite{Donnay:2016ejv} and we will omit it in the following.} of the advanced time $v$ and the angles $z^A$, and assuming the gauge fixing conditions 
\begin{equation}\label{metricGaugeFix}
 g_{\rho \rho }=0\,, \quad g_{v\rho }=1\,, \quad g_{A\rho }=0\,,
\end{equation}
then there exists a set of asymptotic diffeomorphisms preserving~\eqref{NHmetric} generated by an infinite-dimensional algebra that includes both supertranslations and superrotations. These are diffeomorphisms 
\begin{equation}
 \chi = \chi^v\partial_v + \chi^A\partial_A + \chi^{\rho }\partial _{\rho }\,,
\end{equation}
of the form
\begin{equation}\label{chiDiffeos}
\chi^v = f\,, \quad\chi^A=Y^A -\partial_Bf\int ^{\rho } d\rho' g^{AB} \,, \quad\chi^{\rho} = -\rho \partial_vf+\partial_Af \int^{\rho } d\rho' g^{AB}g_{vB}\,,
\end{equation}
where $f$ and $Y^A$ are $\rho$-independent functions whose $v$-dependence is constrained by
\begin{equation} \label{chiConstraint}
\partial_v Y^A = 0\,, \quad \kappa \partial_v f + \partial^2_v f = 0 \,.
\end{equation}
These last two equations follow from demanding the leading terms of $\chi$ not to depend on the fields, and from taking the surface gravity, $\kappa$, to be constant. The diffeomorphisms~\eqref{chiDiffeos} subject to~\eqref{chiConstraint} have been shown to give rise to an infinite-dimensional algebra consisting of two copies of the Virasoro algebra generated by $Y^A$ (superrotations) and two Abelian current algebras generated by $f$ (supertranslations). 
For non-extremal black holes ($\kappa\neq 0$), the time-independent part of $f$ can be interpreted as a supertranslation in the retarded time $v$ of the future horizon $\hor^+$. Its time-dependent part can be thought of as a superdilation in $v$ or, alternatively, as a supertranslation in the affine parameter $\lambda = e^{-\kappa v}$ along the event horizon. For extremal black holes ($\kappa=0$), the roles of supertranslations and superdilations are interchanged.

The diffeomorphisms~\eqref{chiDiffeos} subject to the constraints~\eqref{chiConstraint} preserve the generic form of the metric~\eqref{NHmetric}, but change the functions $\kappa\,, \theta_A\,,\Omega_{AB}$ as follows
\begin{equation}\label{deltachis}
\badat{3}
&\delta_{\chi }\kappa = 0\,,\\
&\delta_{\chi }\theta_{A} = \mathcal{L}_{\chi} \theta_A + f\partial_v\theta_A-2\kappa \partial_Af- 2\partial_v\partial_Af+\Omega^{BC}\partial_v\Omega_{AB}D_C f \,,\\
&\delta_{\chi }\Omega_{AB} = f\partial_v\Omega_{AB}+\mathcal{L}_{\chi }\Omega_{AB}\,.
\eadat
\end{equation}

We will now discuss how the transient gravitational shockwave of HPS~\cite{Hawking:2016sgy} and its deformation of the horizon can be interpreted as the change~\eqref{deltachis} in the near-horizon metric from the bald Schwarzschild black hole~\eqref{baldSchw} to the supertranslated one~\eqref{supertranslatedSchw}. This relates the BMS supertranslation at null infinity to the horizon supertranslations and superroations~\eqref{chiDiffeos}-\eqref{chiConstraint}.

To make contact with the asymptotic symmetry analysis of~\cite{Donnay:2015abr,Donnay:2016ejv}\footnote{The infinite-dimensional symmetries at the horizon have been also discussed in references \cite{Blau:2015nee, Afshar:2016wfy, Mao:2016pwq, Penna:2015gza, Penna:2017bdn, Lust:2017gez, Bousso:2017rsx, Eling:2016xlx, Chandrasekaran:2018aop}; see also references therein and thereof.} we can write the near-horizon metric of the supertranslated~\eqref{supertranslatedSchw} Schwarzschild black holes in the form~\eqref{NHmetric} by changing coordinates $\rho = r-(r_+)_f$ and expanding~\eqref{supertranslatedSchw} near the horizon. To leading order in $\rho$, this gives
\begin{equation}\label{supertranslatedSchwNH}
ds^2=  - \frac{1}{r_{+}}\rho dv^2 +2dvd\rho -\frac{2}{r_{+}}\rho D_Af dz^Adv +  ( r_{+}^2 \gamma_{AB}+ 2r_{+}D_AD_Bf )dz^Adz^B + \dots\,, 
\end{equation}
where we used $d\rho = dr - ({1}/{2})D_AD^2fdz^A $. 
From this one can read off the $\cO(\rho^0)$ and $\mathcal{O}(\rho)$ contributions to the metric components induced at the horizon of the supertranslated Schwarzschild black hole, namely
\begin{equation}\label{SchwKappaThetaOmega}
\kappa = \frac{1}{2r_{+}} \,, \quad \theta_A=-2\kappa  D_Af\,, \quad  \Omega_{AB} = r_{+}^2 \gamma_{AB} + 2r_{+} D_AD_Bf \,.
\end{equation}
The corresponding metric functions for the bald Schwarzschild geometry~\eqref{baldSchw} are obtained by setting $f=0$ in~\eqref{SchwKappaThetaOmega}.
Asking~\eqref{SchwKappaThetaOmega} to be generated by acting with~\eqref{deltachis} on the geometry of the unperturbed Schwarzschild horizon, we find
\begin{equation}
\delta_{\chi} \theta_A = -2\kappa D_A f\,, \quad \delta_{\chi}\Omega_{AB} = 2r_{+} D_AD_Bf \ ,
\end{equation}
which corresponds to a horizon supertranslation composed with a horizon superrotation, the latter given by\footnote{Notice that we are not assuming the vector $Y^A$ to be a holomorphic function here.}
\begin{equation}\label{YSchw}
Y_A = \frac{1}{r_+ } D_A f\,.
\end{equation}
Here, $f$ is the HPS supertranslation at null infinity $\scri^+$ which turns out to coincide with the horizon supertranslation at $\hor^+$.
This hence shows that the disturbance produced by the gravitational shockwave, which from null infinity is seen as the action of a pure BMS supertranslation on the Schwarzschild geometry, is seen by the near-horizon observer as a supertranslation $f$ together with an induced superrotation $Y^A$ given by~\eqref{YSchw}. 

We can now compute the conserved charges at the horizon associated to the horizon supertranslation and superrotation symmetries.
For spacetimes of the form~\eqref{NHmetric} these charges have been constructed in~\cite{Donnay:2015abr,Donnay:2016ejv} using the covariant formalism~\cite{Barnich:2001jy}. The horizon superrotation charge for the perturbed Schwarzschild black hole~\eqref{supertranslatedSchwNH} is 
\begin{equation}\label{Qhorizon}
\badat{2}
Q_{Y} = \frac{1}{8\pi }\int d^2z \sqrt{\gamma} \ Y^A\theta_A\Omega =\frac{M}{8\pi }\int d^2z\sqrt{\gamma}\ Y^A\partial_A f \,,
\eadat
\end{equation}
where the integration is over the constant $v$ section of the horizon $\cH^+$ and $\Omega_{AB}= \Omega\,\gamma_{AB}$ so that $\Omega=4M^2+\cO(f)$.
Note that the functional form of~\eqref{Qhorizon} is the same\footnote{There is an extra overall factor $-3$ when directly comparing with the expression for $Q_Y^{\text{HPS}}$. These two charges are quantities defined by integrating over different 2-surfaces.} as the one of HPS given in~\eqref{QHPS}. The additional horizon superrotations induce a supertranslation charge contribution
\begin{equation}
\delta Q_{T}= \frac{\kappa}{8\pi }\int d^2z \ f\ \delta (\sqrt{\det \Omega_{AB}}) \,, \label{Qst}
\end{equation}
which is absent at null infinity $\scri^-$.
The physical interpretation of the zero-mode of~\eqref{Qst} is clear: it encodes the variation of the entropy (times the temperature) due to the transient shockwave, namely
\begin{equation}
\delta Q_{T_{|f=1}}= \frac{\kappa}{2\pi} \frac{\delta \mathcal{A}}{4} , \label{Qnosss}
\end{equation}
where $\delta \mathcal{A}$ is the variation of the horizon area in Planck units\footnote{We use units where Newton's constant $G=1$.}. That is, $\delta Q_{T_{|f=1}}=T_H \delta S$, with $S$ being the Bekenstein-Hawking entropy and $T_H$ being the Hawking temperature.

This concludes our discussion of the black hole horizon memory effect for Schwarzschild black holes. 
In the following, we turn on gauge fields which will turn out to act non-trivially on the the horizon superrotation charge. To do so, we first need to extend the asymptotic symmetry analysis of~\cite{Donnay:2015abr,Donnay:2016ejv} to the Einstein-Maxwell theory and then generalize the discussion of the previous sections to Reissner-Nordstr\"{o}m black holes.

\section{Horizon symmetries for Einstein-Maxwell}\label{sec:EinsteinMaxwell}

The near-horizon geometry of a four-dimensional charged black hole takes the same convenient form in Gaussian null coordinates as that of its uncharged counterpart, namely~\eqref{NHmetric}, which we repeat here for convenience:
\begin{equation}\label{RNmetricBdyCond}
\badat{3}
g_{vv}&=-2 \rho \,\kappa +\cO(\rho^2)\,, \\
g_{vA}&= \rho \,\theta_A(z^B) +\cO(\rho^2)\,, \\
g_{AB}&=\Omega_{AB}(z^C)+\rho \,\lambda_{AB}(z^C)+\cO(\rho^2)\,,
\eadat
\end{equation} 
and we assume the following gauge fixing conditions for the metric 
\begin{equation}\label{RNmetricGaugeFix}
 g_{\rho \rho }=0\,, \quad g_{v\rho }=1\,, \quad g_{A\rho }=0\,.
\end{equation}
As in~\cite{Donnay:2015abr}, the functions $\theta_A$ and $\Omega_{AB}$ depend on $z^A$ but are taken to be independent of the advanced time $v$; this accommodates the case of isolated horizons studied here\footnote{One may relax this assumption, but then one has to treat subtleties regarding the integrability of the charges; see \cite{Donnay:2016ejv}.}. The asymptotic boundary conditions for the Maxwell field are
\begin{equation}
\label{RNABdyCond}
\badat{2}
&A_v=A_v^{(0)}+\rho \, A_v^{(1)}(v,z^A)+\rd O(\rho^2)\,,\\
&A_B= A_B^{(0)}(z^A)+\rho \, A_B^{(1)}(v,z^A)+\rd O(\rho^2)\,,
\eadat
\end{equation}
and we choose the radial gauge condition
\begin{equation}\label{AGaugeFix}
 A_\rho=0 \,.
\end{equation}
The Coulombian potential at the horizon, $A_v^{(0)}$, is taken to be a fixed constant, $A_B^{(0)}$ is assumed to depend only on $z^A$, while $A_v^{(1)}$ and $A_B^{(1)}$ are arbitrary functions of $z^A$ and $v$. These conditions are analogous to those considered for Einstein-Maxwell theory at null~\cite{Barnich:2013sxa, Barnich:2015jua} and spatial~\cite{Henneaux:2018gfi} infinity. They are slightly more general than the horizon conditions considered in~\cite{Mao:2016pwq}, and suffice to discuss physically interesting solutions such as Kerr-Newman black holes. We now study the horizon symmetries for metrics and gauge fields obeying~\eqref{RNmetricBdyCond}-\eqref{AGaugeFix}. Depending on what is more convenient in specific examples, we will either consider the angular coordinates $z^A$ to be parametrized by complex variables $(z,\bar{z})$ or by standard polar variables $(\theta,\phi)$.

A set of field transformations at the horizon is given by
\begin{equation} \label{gauget}
\delta _{(\chi, \ep)}g_{\mu \nu}=\cL_\chi g_{\mu \nu}\,, \quad  \delta _{(\chi, \ep)}A_\mu=\cL_\chi A_\mu+\p_\mu \epsilon,
\end{equation}
where the vector field $\chi=\chi^\mu \p_\mu$ generates diffeomorphisms while $\ep$ is the gauge parameter. These transformations represent symmetries at the horizon if they respect the asymptotic form~\eqref{RNmetricBdyCond}-\eqref{RNABdyCond}. The gauge fixing conditions~\eqref{RNmetricGaugeFix} and~\eqref{AGaugeFix} imply
\begin{equation}\label{LGaugeFix}
\cL_\chi g_{\rho \rho}=0 \,, \quad \cL_\chi g_{v \rho }=0 \,, \quad \cL_\chi g_{\rho A}=0 \,, \quad  \cL_\chi A_\rho+\p_\rho \epsilon=0\,,
\end{equation}
 which yield the following form for $\chi^{\mu }$ and $\epsilon $ 
\begin{equation}
\label{solexact}
\badat{3}
&\chi^{v} = f(v,z^A), \\
&\chi^{\rho } = Z(v,z^A)- \rho \p_{v}f + \p_{A}f \int^\rho_0 d\rho' g^{AB}g_{vB},  \\
&\chi^{A} = Y^A(v,z^A) -\p_{B}f \int^\rho_0 d\rho' g^{AB},\\
&\ep=\ep^{(0)}(v,z^A)- \int^\rho_0 d\rho' A_B \partial_\rho \chi^B ,
\eadat
\end{equation}
where $f$, $Z$, $Y^A$ and $\ep^{(0)}$ are arbitrary functions of $v$ and $z^A$ that do not depend on $\rho$. Demanding that the leading piece of the vector field $\chi$ only depends on the coordinates but not on the fields (i.e. the arbitrary functions appearing in the metric and gauge field) leads to $Z=0$ and $\partial_v Y^A=0$~\cite{Donnay:2016ejv}. Implementing the boundary conditions for the remaining metric components, namely
\begin{equation}
 \cL_\chi g_{vv}=\cO(\rho^2) \,, \quad \cL_\chi g_{vA }= \cO(\rho^2)\,, \quad \cL_\chi g_{AB}= \cO(\rho)\,, 
\end{equation}
yields the following components of the diffeomorphism generating vector field
\begin{equation}
\badat{3}
\label{chisNonExt}
&\chi^{v} = f(v,z^A), \\
&\chi^{\rho } =-\p_v f(v,z^A) \rho + \frac{\rho^2}{2 \Omega}\theta_A(z^B) \partial^{A} f(v,z^A) + O(\rho^3),  \\
&\chi^{A} = Y^A(z^B) -\frac{\rho}{\Omega} \partial^A f(v,z^A)  + \frac{\rho^2}{2 \Omega^2} \lambda^{AB}(z^C) \p_{B} f(v,z^A)  + O(\rho^3),
\eadat
\end{equation}
where we used the conformal gauge $\Omega_{AB}=\Omega \gamma_{AB}$, which implies that the vector $Y^A$ is a conformal Killing vector on the 2-sphere.
For the gauge field, the boundary conditions~\eqref{RNABdyCond} imply
\begin{equation}
\cL_\chi A_v+\p_v \epsilon=\rd O(\rho)\,, \quad \cL_\chi A_B+\p_B \epsilon=\rd O(1)\,,
\end{equation}
yielding
\begin{equation}
\epsilon^{(0)}(v,z^A)=U(z^A)-f(v,z^A) A_v^{(0)} \,,
\end{equation}
where $U$ is an arbitrary function of the angular coordinates $z^A$. This yields the gauge parameter
\begin{equation}\label{epsilon}
\ep=U(z^A)-f(v,z^A) A_v^{(0)}  + \rho\ {\Omega}^{-1}{\partial^B f(v,z^A)}  A_B^{(0)}(z^A) +\rd O(\rho^2)\,.
\end{equation}
Thus we find that the transformations~\eqref{gauget} for the diffeomorphism vector field~\eqref{chisNonExt} and the gauge parameter~\eqref{epsilon} generate the horizon symmetries. The variations of the functions $\kappa,\theta_A,\Omega_{AB}$ of the metric and the angular part of the gauge field $A_v,A_B$ are 
\begin{equation}
\badat{5} \label{deltaRN}
&\delta_{(\chi,\ep)} \kappa =0= \kappa \p_v f + \p_v^2 f\,,  \\ 
&\delta_{(\chi,\ep)} \theta_A =\cL_Y \theta_{A}+ f \partial_v\theta_A -2\kappa  \partial_A f -2\partial_v\partial_A f +\Omega^{BC} \partial_v\Omega_{AB} \partial_C f \,,\\
&\delta_{(\chi,\ep)}  \Omega_{AB} = f\partial_v\Omega_{AB} + {\mathcal L}_{Y}\Omega_{AB}\,,\\
&\delta_{(\chi,\ep)} A_v^{(0)}=0\,,\\
&\delta_{(\chi,\ep)} A_B^{(0)}=Y^C \p_C A_B^{(0)}+A_C^{(0)} \p_B Y^C+ \p_B U\,.
\eadat
\end{equation}
These variations generate a Lie algebra. If the gauge parameters $\chi$ and $\epsilon$ depended only on the spacetime coordinates but not on the fields, the Lie product
\begin{equation}
[\delta _{(\chi_1, \ep_1)},\delta _{(\chi_2, \ep_2)}](g_{\mu \nu}, A_\mu)=\delta _{(\hat \chi, \hat \ep)}(g_{\mu \nu}, A_\mu),
\end{equation}
would take a simple form with $\hat \chi=[\chi_1,\chi_2]$ and $\hat \ep=\chi_1^\mu \p_\mu \ep_2-\chi_2^\mu \p_\mu \ep_1$; that is, the Lie bracket would be
\begin{equation}
[(\chi_1, \ep_1),(\chi_2, \ep_2)]=(\hat \chi, \hat \ep).
\end{equation}
However, when the gauge parameters do depend on the fields, as in (\ref{chisNonExt})-(\ref{epsilon}), one needs to resort to the modified Lie bracket~\cite{Barnich:2010eb}
\begin{equation}
[(\chi_1, \ep_1),(\chi_2, \ep_2)]_M=(\hat \chi, \hat \ep)\,,
\end{equation}
where now 
\begin{equation}
\badat{2}
&\hat \chi=[\chi_1,\chi_2]+\delta _{(\chi_1, \ep_1)}\chi_2-\delta _{(\chi_2, \ep_2)}\chi_1\,, \\
&\hat \ep=\chi_1^\mu \p_\mu \ep_2-\chi_2^\mu \p_\mu \ep_1+\delta _{(\chi_1\,, \ep_1)}\ep_2-\delta _{(\chi_2, \ep_2)}\ep_1\,.
\eadat
\end{equation}
With this modified bracket, one finds that the parameters~\eqref{chisNonExt} and~\eqref{epsilon} of the residual gauge symmetries form a representation of the infinite-dimensional Lie algebra which can be expressed as
\begin{equation}\label{closureNonExt}
  [(f_1,Y_1^A,U_1),(f_2,Y_2^A,U_2)]=(\hat f,\hat Y^A, \hat U)\,,
\end{equation}
with
\begin{equation} \label{modifiedLie}
\badat{3}
&\hat f=f_1 \p_{v} f_2+Y_1^A \p_{A} f_2-(1\leftrightarrow 2)\,,\\
& \hat Y^A=Y_1^B  \p_{B}  Y_2^A -(1\leftrightarrow 2)\,,\\
&\hat U=Y_1^A \p_A U_2-(1\leftrightarrow 2)\,.
\eadat
\end{equation}
From this point on, the asymptotic symmetry analysis has to be treated separately for non-extremal and extremal horizons.

\subsection{Non-extremal horizons}\label{subsec:neRN}

For isolated non-extremal horizons ($\kappa=\text{const}\neq0$), the first equation in~\eqref{deltaRN} yields the linear equation 
\begin{equation}\label{fNonExtConstraint}
0 = \kappa \partial_v  f + \partial_v^2f \,, 
\end{equation}
which has a solution of the form
\begin{equation}\label{fNonExt}
f(v,z^A) = T(z^A) + e^{-\kappa  v}\ X(z^A) \,.
\end{equation}
We see from this that there are two distinct contributions to the supertranslation generator which have different physical interpretations. The first term in~\eqref{fNonExt} generates supertranslation charge at the horizon. The exponential decay in advanced time in the second term in~\eqref{fNonExt} resembles the so-called horizon redshift effect~\cite{Lucietti:2012xr}, where the energy of a photon moving tangential to the horizon undergoes a redshift proportional to $e^{-\kappa v}$. The wave analog of this effect is important in proving the linear stability of Schwarzschild and non-extreme Kerr spacetimes under scalar perturbations~\cite{Dafermos:2008en,Dafermos:2010hd}. In terms of these two different contributions the algebra~\eqref{closureNonExt} closes with
\begin{equation}\label{TXNonExt}
\badat{2}
&\hat{T}=Y_1^A \p_{A} T_2-(1\leftrightarrow 2)\,,\\
&\hat{X}=Y_1^A  \p_{A}  X_2 -\kappa T_1 X_2 -(1\leftrightarrow 2)\,.
\eadat
\end{equation}

Expanding the superrotation, supertranslation and electromagnetic charge generators in modes 
\begin{equation}\label{ModeExpansionNonExt}
 \badat{3}
 &T(z,\bar{z})=\sum_{m,n}T_{(m,n)}z^m\bar{z}^n\,,\quad X(z,\bar{z})=\sum_{m,n}X_{(m,n)}z^m\bar{z}^n\,,\\
 &Y^z(z)=\sum_{n}z^nY_{n}\,,\quad Y^{\bar{z}}(\bar{z})=\sum_{n}\bar{z}^n\bar{Y}_{n}\,,\\
 &U (z,\bar{z})=\sum_{m,n}U_{(m,n)}z^m\bar{z}^n\,,
 \eadat
\end{equation}
where we used complex coordinates for the angular variables $z^A=( z,\bar{z} )$ and $m,n\in\mathbb{Z}$, 
the algebra~\eqref{closureNonExt} becomes
\begin{equation}\label{ModeAlgebraNonExt}
\badat{6}
&[ Y_m , Y_n ] = (m-n) Y_{m+n}, \\  
&[ \bar Y_m , \bar Y_n ] =  (m-n) \bar Y_{m+n},\\
&[ Y_k , T_{(m,n)} ] =  - m T_{(m+k,n)},\\
&[\bar Y_k , T_{(m,n)} ] =  - n T_{(m,n+k)},\\
&[ Y_k , X_{(m,n)} ] =  - m X_{(m+k,n)},\\
&[\bar Y_k , X_{(m,n)} ] =  - n X_{(m,n+k)},\\
&[ X_{(k,l)} ,T_{(m,n)}] = \kappa X_{(m+k,n+l)}, \\
&[ Y_k , U_{(m,n)} ] =  - m U_{(m+k,n)},\\
&[ \bar Y_k , U_{(m,n)} ] =  - n U_{(m+k,n)},
\eadat
\end{equation}
with the remaining commutators being zero. This algebra contains three sets of supertranslations currents, generated by $T_{(m,n)}$, $X_{(m,n)}$ and $U_{(m,n)}$, and two sets of Virasoro (Witt) modes $Y_n , \bar{Y}_n$ which are in semi-direct sum with the supertranslations. The algebra contains ideals, generated by $X_{(m,n)}$ and $U_{(m,n)}$. The supertranslation charge generator $T_{(m,n)}$ does not commute with $X_{(m,n)}$ but does commute with the generator of electromagnetic charge $U_{(m,n)}$. {The supertranslation zero-mode $T_{(0,0)}$ corresponds to the Killing vector associated to rigid translations in the advanced time $v$, and consequently is associated to a notion of energy. A large set of generators $Y_m$, $\bar{Y}_m$, $U_{(m,n)}$, and $T_{(m,n)}$ commutes with this energy operator; it is thus natural to refer to them as soft horizon hairs. The generators $X_{(m,n)}$, in contrast, behave under the action of $T_{(0,0)}$ as an expansion: $[X_{(m,n)} , T_{(0,0)}] = \kappa X_{(m,n)}$.} Hence, one may wonder about the existence of an additional conformal symmetry. However, as we will show next, $X$ does not appear in the conserved charge; and so we can conclude that it is pure gauge.

Diffeomorphisms and gauge symmetry transformations generated by the modes~\eqref{ModeAlgebraNonExt} have an associated set of conserved charges. The latter can be computed using again the method of~\cite{Barnich:2001jy}, including now the gauge field contribution. We find\footnote{The convention for the action used here is given in~\eqref{EMAction} below.}
\begin{equation}
Q[T,Y^A,U]=\frac{1}{16\pi } \int d^2z \sqrt{\gamma} \Omega \left( 2T\kappa - Y^A \theta_A -4 U A_v^{(1)} - 4 A_B^{(0)} Y^B A_v^{(1)}  \right)\,,
\label{QEM}
\end{equation}
where there is indeed no contribution from $X$.
The first three terms in~\eqref{QEM} correspond to the horizon charges computed in \cite{Donnay:2016ejv}, while the fourth term is purely of electric origin, and the last term mixes the electromagnetic field with the superrotation vector field contribution. The charge~\eqref{QEM} evaluated on the modes~\eqref{ModeAlgebraNonExt} obeys the algebra
\begin{equation}\label{ChargeAlebraNonExt}
\badat{6}
&\{ \rd Y_m , \rd Y_n \} = (m-n) \rd Y_{m+n}, \\  
&\{ \bar {\rd Y_m} , \bar{\rd  Y_n} \} =  (m-n) \bar {\rd Y}_{m+n},\\
&\{ \rd  Y_k ,\rd  T_{(m,n)} \} =  - m \rd T_{(m+k,n)},\\
&\{ \bar {\rd Y_k} , \rd T_{(m,n)} \} =  - n \rd T_{(m,n+k)},\\
&\{ \rd Y_k , \rd U_{(m,n)} \} =  - m \rd U_{(m+k,n)},\\
&\{ \bar {\rd Y_k} , \rd U_{(m,n)} \} =  - n \rd U_{(m,n+k)},
\eadat
\end{equation}
where $\mathcal{Y}_{n}$, $\bar{\mathcal{Y}}_{n}$, $\mathcal{T}_{(m,n)}$, and $\mathcal{U}_{(m,n)}$ are the charges associated to the modes ${Y}_{n}$, $\bar{{Y}}_{n}$, ${T}_{(m,n)}$, and ${U}_{(m,n)}$, respectively. The brackets in~\eqref{ChargeAlebraNonExt} are defined as in~\cite{Donnay:2016ejv}. The charge~\eqref{QEM} and the algebra~\eqref{ChargeAlebraNonExt} generalize the results of~\cite{Donnay:2015abr} to include $U(1)$ gauge fields. The generalization to the case of $N$ Abelian gauge fields is straightforward by considering $\mathcal{U}_{(m,n)}\to \mathcal{U}^I_{(m,n)}$ with $I=1,2,... N$ with $\{\mathcal{U}_{(m,n)}^I, \mathcal{U}_{(k,l)}^J\}=0$.

In order to investigate the physical meaning of the charge~\eqref{QEM}, we can consider the static Reissner-Nordstr\"{o}m solution to Einstein-Maxwell theory. The action of the theory is
\begin{equation}\label{EMAction}
S=\frac{1}{16\pi } \int d^4x \sqrt{g}\left(R-F_{\mu \nu}F^{\mu \nu}\right)\,.
\end{equation}

The dyonic Reissner-Nordstr\"{o}m solution in advanced Eddington-Finkelstein coordinates $(v,r,\theta,\phi)$ and a suitable gauge\footnote{This amounts to performing a gauge transformation with parameter $d\lambda = -q\Delta^{-1}rdr$ on the standard form of the gauge field.} is given by the following metric and gauge field 
\begin{equation}\label{baldRN}
\badat{2}
 ds^2&=-\frac{\Delta}{r^2} dv^2 +2 dv dr +r^2 (d\theta^2+\sin^2\theta d\phi^2)\,,\\
A&=-\frac{q}{r}dv-p(\cos\theta{-k}) d\phi \,, 
\eadat
\end{equation}
where $\Delta=r^2+e^2-2Mr$ with $e^2={q^2+p^2}$, where $q$ and $p$ are the electric and magnetic charges, respectively. The outer horizon is located at $r_+=M+\sqrt{M^2-e^2}$. The constant $k$ appearing in the gauge field is in principle arbitrary and can be changed by a gauge transformation. However, for $k=\pm 1$ the solution exhibits a special property: Provided $p\neq 0 $, the gauge field configuration above is singular on the axis $\theta =0 $, where a Dirac string exists. Then, following standard practice, one can choose different gauges in each hemisphere, in such a way that either the north pole or the south pole is singularity free. This is achieved by choosing $k=1$ or $k=-1$, respectively.

After the coordinate transformation $\rho = r-r_+$, the (outer) horizon is located at $\rho=0$ 
and the metric and the gauge field take a form suitable for comparison with the asymptotic symmetry analysis of the previous section. Expanding the metric near the horizon and comparing to~\eqref{RNmetricBdyCond}, we can read off:
\begin{equation}\label{horfuncRN}
 \kappa=\frac{(r_+-M)}{r_+^2}\,, \quad \theta_A=0\,, \quad \Omega_{\theta\theta}=r_+^2\,, \quad \Omega_{\phi \phi}=r_+^2\sin^2\theta\,,\quad \Omega_{\theta \phi}=0\,.  
\end{equation}
The expansion of the gauge field near $\rho=0$ yields 
\begin{equation}\label{horARN}
 {A_\rho=0\,, \quad A_v^{(0)}=-\frac{q}{r_+}\,,  \quad A_v^{(1)}=\frac{q}{r_+^2}\,, \quad A_B^{(0)}=-p(\cos\theta {-k})}\delta^{\phi }_B\,.
\end{equation}
The surface charge at the horizon for this static configuration with~\eqref{horfuncRN} and~\eqref{horARN} yields
\begin{equation}\label{QhorNonExtRN}
 Q[T,Y^{\phi},U]=\frac{1}{16\pi} \int d\theta d\phi \sin\theta \left(2{(r_+-M)} T -4{q}U+4{pq}(\cos\theta {-k}) Y^{\phi}\right)\,,
\end{equation}
where the range of integration over the constant $v$ section of the horizon has to be chosen such that the singularities of $A$ at $\theta =0$ and $\theta =\pi$ are avoided. This yields the zero-modes\footnote{The charge $\mathcal{Y}_{(0,0)}$, associated to the rigid translations $\partial_{\phi }$, in terms of the complex variables $z,\bar{z}$ is given by the charge $\mathcal{Y}_{0}-\bar{\mathcal{Y}}_{0}$.} 
\begin{equation}\label{ZeroModesNonExt}
\badat{3}
 &\mathcal{T}_{(0,0)}&\equiv& Q[1,0,0]= \kappa \frac{r_+^2 }{2}\,,\\
& \mathcal{U}_{(0,0)}&\equiv& Q[0,0,1]=-q\,,\\
& \mathcal{Y}_{(0,0)}&\equiv& Q[0,1,0]={0}\,.
\eadat
\end{equation}
These three different contributions have the following interpretation: The first one, $\mathcal{T}_{(0,0)}$, has a simple interpretation in the context of black hole thermodynamics, as it gives the product between the Hawking temperature $T_H=\kappa /(2\pi )$ and the Bekenstein-Hawking entropy $S=\pi r_+^2$. The second contribution, $\mathcal{U}_{(0,0)}$, corresponds to the electric charge of the black hole. Finally, the third contribution, $\mathcal{Y}_{(0,0)}$, gives the angular momentum of the black hole~\cite{Donnay:2015abr}. In the case of the static Reissner-Nordstr\"{o}m solution, this gives zero\footnote{The computation of this charge yields $\mathcal{Y}_{(0,0)}=-({qp}/{2})(\int_{0}^{{\pi }/{2}} d\theta \sin \theta (\cos\theta -1) + \int_{{\pi }/{2}}^{\pi } d\theta \sin \theta (\cos\theta +1)) =0$. Notice that had we performed the integral over the range $\theta \in [0,\pi ]$ without changing the gauge in each hemisphere, we would have rather obtained the result $\mathcal{Y}_{(0,0)}=-pqk$. This non-vanishing result comes from integrating a singular configuration of $A$: it can be interpreted as the contribution from the Dirac string.} which is consistent with the fact that the contribution of the electromagnetic field of the dyonic black hole to the total angular momentum is zero \cite{Garfinkle:1990zx,Bunster:2007sn}.

\subsection{Extremal horizons}\label{subsec:eRN}

In the above discussion of the horizon symmetries of the Reissner-Nordstr\"{o}m black hole we assumed $\kappa=\text{const}\neq 0$. The extremal limit, corresponding to $\kappa=0$, has to be treated separately. In particular, in this case equation~\eqref{fNonExtConstraint} becomes
\begin{equation}
\p_v^2 f=0,
\end{equation}
whose solution
\begin{equation}\label{fExt}
f=T(z,\bar{z})+v\ X(z,\bar{z}) , 
\end{equation} 
contains a linearly growing term in advanced time $v$ rather than an exponentially decaying one as in~\eqref{fNonExt}. This modifies the condition for closure of the algebra from~\eqref{TXNonExt} to
\begin{equation}\label{TXExt}
\badat{3}
&\hat T=T_1 X_2+Y_1^A \p_{A} T_2-(1\leftrightarrow 2),\\
&\hat X=Y_1^A  \p_{A}  X_2 -(1\leftrightarrow 2)\,.
\eadat
\end{equation}
Expanding in modes we find the following  algebra
\begin{equation}\label{ModeAlgebraExt}
\badat{6}
&[ Y_m , Y_n ] = (m-n) Y_{m+n}, \\  
&[ \bar Y_m , \bar Y_n ] =  (m-n) \bar Y_{m+n},\\
&[ Y_k , T_{(m,n)} ] =  - m T_{(m+k,n)},\\
&[\bar Y_k , T_{(m,n)} ] =  - n T_{(m,n+k)},\\
&[ Y_k , X_{(m,n)} ] =  - m X_{(m+k,n)},\\
&[\bar Y_k , X_{(m,n)} ] =  - n X_{(m,n+k)},\\
&[ X_{(k,l)} ,T_{(m,n)}] = T_{(m+k,n+l)}, \\
&[ Y_k , U_{(m,n)} ] =  - m U_{(m+k,n)},\\
&[ \bar Y_k , U_{(m,n)} ] =  - n U_{(m+k,n)},
\eadat
\end{equation}
with the other commutators being zero. It is interesting to compare~\eqref{ModeAlgebraExt} with~\eqref{ModeAlgebraNonExt}. From the non-extremal algebra~\eqref{ModeAlgebraNonExt} one would naively expect $X_{(k,l)}$ and $T_{(m,n)}$ to commute when $\kappa=0$. However,~\eqref{ModeAlgebraExt} shows that this is clearly not the case; the limit $\kappa\to 0$ is not continuous. Further comparison of the commutator $[X_{(k,l)},T_{(m,n)}]$ reveals that the roles of $X_{(m+k,n+l)}$ and $T_{(m+k,n+l)}$ are interchanged between non-extremal and extremal horizons.

The set of conserved charges associated to~\eqref{ModeAlgebraExt} is 
\begin{equation}\label{QhorExtRN}
Q[X,Y^{A},U]=\frac{1}{16\pi } \int dz d\bar{z} \sqrt{\gamma}\Omega   \Bigg( 2X - Y^A \theta_A -4U A_v^{(1)} - 4A_B^{(0)} Y^B A_v^{(1)} \Bigg) .
\end{equation}
Notice that in contrast to~\eqref{QhorNonExtRN} there is no dependence on $T$ in~\eqref{QhorExtRN}. From this we conclude that the modes associated to $T$ become pure gauge in the extremal case.

The algebra generated by these charges is the same as the one obeyed by the vector fields~\eqref{ModeAlgebraExt}, namely
\begin{equation}\label{ChargeAlebraExt}
\badat{6}
&\{ \rd Y_m , \rd Y_n \} = (m-n) \rd Y_{m+n}, \\  
&\{ \bar {\rd Y_m} , \bar {\rd Y_n } \} =  (m-n) \rd {\bar Y}_{m+n},\\
&\{ \rd Y_k , \rd X_{(m,n)} \} =  - m \rd X_{(m+k,n)},\\
&\{ \bar {\rd Y_k} , \rd X_{(m,n)} \} =  - n \rd X_{(m,n+k)}\\
&\{ \rd Y_k , \rd U_{(m,n)} \} =  - m \rd U_{(m+k,n)},\\
&\{ \bar {\rd Y_k} , \rd U_{(m,n)} \} =  - n \rd U_{(m,n+k)}.
\eadat
\end{equation}
This is similar to the non-extremal case~\eqref{ChargeAlebraNonExt}: The algebra contains two copies of the Virasoro algebra, generated by $\mathcal{Y}_n$ and $\bar{\mathcal{Y}}_n$, and two affine currents, generated now by $\mathcal{X}_{(m,n)}$ and $\mathcal{U}_{(m,n)}$.

Repeating the steps of Section~\ref{subsec:neRN} for extremal Reissner-Nordstr\"{o}m black holes, which have $|e|=M$, yields the following metric functions
\begin{equation}\label{horfunceRN}
 \kappa=0\,, \quad \theta_A=0\,, \quad \Omega_{\theta\theta}=r_+^2\,, \quad \Omega_{\phi \phi}=r_+^2 \sin^2\theta\,,\quad \Omega_{\theta \phi}=0\,,
\end{equation}
and gauge field components
\begin{equation}
 {A_\rho=0\,,  \quad A_v^{(0)}=-\frac{q}{r_+}\,,  \quad A_v^{(1)}=\frac{q}{r_+^2}\,, \quad A_B=-p(\cos\theta {-k})}\delta^\phi_B\,.
\end{equation}
Evaluating the surface charge at the horizon on this extremal configuration gives
\begin{equation}
 Q[X,Y^{\phi},U]=\frac{1}{16\pi} \int d\theta d\phi \sin\theta r_+^2\left(2X-4\frac{q}{r_+^2}U+4\frac{pq}{r_+^2}(\cos\theta-k) Y^{\phi }\right)\,,
\end{equation}
whose zero-modes are
\begin{equation}\label{ZeroModesExt}
\badat{3}
& \mathcal{X}_{(0,0)}\equiv Q[1,0,0]=\frac{r_+^2}{2}\,,\\
& \mathcal{U}_{(0,0)}\equiv Q[0,0,1]=-q\,,\\
& \mathcal{Y}_{(0,0)}\equiv Q[0,1,0]={0}\,.
\eadat
\end{equation}
The thermodynamic interpretation of $\mathcal{X}_{(0,0)}$ is slightly different from that of $\mathcal{T}_{(0,0)}$ of the non-extremal case since the Hawking temperature vanishes for extremal black holes. Nevertheless, if we treat the $S^1$ defined by the electromagnetic field as a geometrical fiber as in~\cite{Hartman:2008pb}, we can define a geometric temperature $T_q={1}/({2\pi q})$. The $U(1)$ gauge symmetry gets extended to a Virasoro algebra and the central charge associated to the fiber direction is $c=6q(q^2+p^2)$. This 
yields the black hole entropy $S=({\pi^2}/{3}) c T_q=\pi M^2=\pi r_+^2$ via the Cardy formula. The zero-mode $\mathcal{X}_{(0,0)}$ is then interpreted as the product between the geometric temperature and Bekenstein-Hawking entropy: $qT_q S$. This is analogous to what happens with extremal Kerr black holes in \cite{Donnay:2016ejv}, where the charge $\mathcal{X}_{(0,0)}$ gives the product of the black hole entropy and the geometrical temperature $T_L$ that appears in the Kerr/CFT analysis of the extremal Frolov-Thorne vacuum. As in the non-extremal case, the zero-modes $\mathcal{U}_{(0,0)}$ and $\mathcal{Y}_{(0,0)}$ give, respectively, the electric charge of the black hole and the total angular momentum.

\section{Soft hair on Reissner-Nordstr\"{o}m horizons}\label{sec:MemoryRN}

We are now ready to discuss the horizon memory effect for charged black holes. As for the Schwarzschild geometry in Sections~\ref{sec:HPS}-\ref{sec:MemorySchw} we will first determine the supertranslated Reissner-Nordstr\"{o}m solution obtained from the action of a BMS supertranslation at null infinity\footnote{In reference \cite{Strominger:2017zoo}, a physical process that can be thought of as reciprocal to the one discussed here was considered. There, a null incoming shockwave with asymmetric null charge is sent into an uncharged black hole in such a way that large gauge currents are excited at null infinity.}. We then study whether it is possible to reinterpret this new solution to Einstein-Maxwell theory as the action of the horizon symmetry generators found in Section~\ref{sec:EinsteinMaxwell} on the bald geometry and identify the relation between the symmetry generators at $\scri^-$ and $\hor^+$.

We consider a diffeomorphism generated by an asymptotic BMS vector $\zeta=\zeta^{v}\partial_v+\zeta^{A}\partial_A+\zeta^{r}\partial_r$ at past null infinity $\scri^-$ which preserves the gauge fixing conditions~\eqref{RNmetricGaugeFix} and~\eqref{AGaugeFix} together with the falloffs~\eqref{RNmetricBdyCond} and~\eqref{RNABdyCond} at large radial distance $r$. The gauge fixing requirements translate into the conditions
\begin{equation}\label{LRN}
\badat{3}
& \cL_\zeta g_{rA}=\partial_A \zeta^v+g_{AB} \partial_r \zeta^B=0\,,\\
& \cL_\zeta g_{rr}=2\partial_r \zeta^v=0\,,\\
& \frac{r}{2}g^{AB} \cL_\zeta g_{AB}=r D_A \zeta^A+2\zeta^r=0\,,\\
& \cL_\zeta A_{r}+\p_{r}\epsilon=0\,.
\eadat
\end{equation}
A solution to~\eqref{LRN} is given by
\begin{equation}\label{zetaf}
 \zeta =f \partial_v+\frac{1}{r} D^A f \partial_A-\frac{1}{2} D^2f \partial_r\,, \quad\epsilon=-\frac{1}{r} A_{B} D^B f \,.
\end{equation}
with $\partial_r f=\partial_v f=0$. 
Following~\cite{Hawking:2016sgy}, the asymptotic Killing vector~\eqref{zetaf} extends the asymptotic expansion of the supertranslations on $\scri^-$ to the entire region covered by the advanced Eddington-Finkelstein coordinates, which includes $\mathcal H^+$. The action of~\eqref{zetaf} on the bald Reissner-Nordstr\"{o}m metric~\eqref{baldRN} gives 
\begin{equation}
\badat{3}
& \cL_\zeta g_{vv}=\left(M-\frac{e^2}{r}\right) \frac{D^2f}{r^2}\,,\\
& \cL_\zeta g_{A v}= -D_A \left(\frac{\Delta}{r^2} f +\frac{1}{2} D^2f\right)\,,\\
& \cL_\zeta g_{AB} = -r \gamma_{AB} D^2f +2r D_A D_B f\,,\\
& \cL_\zeta A_{v}+\p_{v}\epsilon= -\frac{1}{2} D^2f  F_{r v}\,, \\
& \cL_\zeta A_{B}+\p_{B}\epsilon=-\frac{q}{r}D_{B} f +\frac{1}{r} F_{A B} D^A f  \,. 
\eadat
\end{equation}
where $F_{rv}=q/r^2$ and $F_{AB}=\epsilon_{AB}\, p \sin \theta$.
The infinitesimally supertranslated Reissner-Nordstr\"{o}m geometry is thus given by
\begin{equation}\label{supertranslatedRN}
\badat{2}
ds^2&=-\left(\frac{\Delta}{r^2}-\left(M-\frac{e^2}{r}\right) \frac{D^2f}{r^2}\right)dv^2+2dv dr-2D_A\left(\frac{\Delta}{r^2} f+\frac{1}{2} D^2 f\right) dv dz^A+\\
& +\left( r^2\gamma_{AB}+2r D_A D_B f-r \gamma_{AB} D^2f\right) dz^A dz^B\,.
\eadat
\end{equation}
The location of the supertranslated (outer) horizon at linear order in $D^2 f$ is
\begin{equation}
 (r_+)_{f}=r_+ +\frac{1}{2} D^2 f\,,
\end{equation}
with $r_+=M+\sqrt{M^2-e^2}$. 

We can now ask whether the spacetime~\eqref{supertranslatedRN} can be obtained by acting on the bald Reissner-Nordstr\"{o}m solution~\eqref{baldRN} with horizon symmetry generators of the type~\eqref{chisNonExt} and~\eqref{epsilon} studied in Section~\ref{sec:EinsteinMaxwell}. To do so, we first need to bring the supertranslated metric~\eqref{supertranslatedRN} to the near-horizon form~\eqref{RNmetricBdyCond}. This is achieved by the following coordinate transformation
\begin{equation}\label{Ortodoncia}
 \rho=r-(r_+)_{f}\,, \quad d\rho=dr-\frac{1}{2} D_A D^2f dz^A\,.
\end{equation}
At order $\mathcal{O}(f)$, this yields the metric functions
\begin{equation}\label{RNKappaThetaOmega}
\badat{4}
 \kappa=\frac{r_+-M}{r_+^2}\,,\,\, 
 \theta_A=-2\kappa D_A f\,,\,\,
 \Omega_{AB}=r_+^2\gamma_{AB}+2r_+ D_A D_B f \,.
 \eadat
\end{equation}
Asking~\eqref{RNKappaThetaOmega} to be generated by the action of~\eqref{deltaRN} on the bald geometry, we find that it is achieved by the horizon superrotation\footnote{Notice that we are not requiring the vector $Y^A$ to be a holomorphic function here.} 
\begin{equation}
 Y_A=\frac{1}{r_+} D_A f\,.\label{DFGHJ}
\end{equation}
Here $f$ is the BMS supertranslation at $\scri^-$ but, as in the case of Schwarzschild, it coincides with the supertranslation at $\hor^+$. Note that this computation is valid also in the extremal case.

Let us now consider the gauge field. In the coordinates~\eqref{Ortodoncia}, the transformed gauge field at the horizon takes the form
\begin{equation}\label{RNA}
A_v + \delta_{(\zeta , \epsilon)}A_v =-\frac{q}{r_+} \,, \quad 
A_{B } + \delta_{(\zeta , \epsilon)}A_{B } =-p \delta^{\phi }_{B}(\cos\theta {-k})-\frac{1}{r_+}(qD_{B} f - F_{A B} D^A f  ) \,.
\end{equation}
This is consistent with the boundary conditions~\eqref{RNABdyCond}, with the field variations $\delta_{(\zeta , \epsilon)}A^{(0)}_v=0$ and $\delta_{(\zeta , \epsilon)}A^{(0)}_B=-(qD_{B} f - F_{A B} D^A f  )/r_+$. Then, from (\ref{deltaRN}) and taking into account (\ref{DFGHJ}) we find the gauge parameter
\begin{equation}
U=-\frac{q}{r_+}f-\frac{1}{r_+}A_B^{(0)}D^Bf\,,\label{DFGHJ2}
\end{equation}
which, as in the case of the metric, yields a non-trivial charge. Therefore, we conclude that~\eqref{RNKappaThetaOmega} and~\eqref{RNA} can indeed be interpreted as the action of the horizon symmetry transformations~\eqref{deltaRN} on the bald Reissner-Nordstr\"{o}m solution.

\section{Conclusions}

Motivated by the problem of establishing a connection between the symmetries emerging in the near-horizon region of black holes and the symmetries in the far asymptotic region, in this paper we studied the memory effect produced at the horizon by an incoming gravitational shockwave. From the point of view of an observer in the asymptotic region, this process was studied in \cite{Hawking:2016sgy}, where it was shown that the shockwave produces a disturbance in the black hole geometry that can be interpreted as a BMS supertranslation at null infinity. Here, we have shown that, from the perspective of an observer hovering close to the event horizon, the shockwave produces not only a supertranslation but also a horizon superrotation. The zero-mode contribution of the horizon superrotation charge is found to take the same form as the one computed by HPS at infinity, while the zero-mode contribution of the horizon supertranslation charge captures the change in the black hole entropy in the process \cite{Donnay:2015abr}. 

We also considered the case of charged black holes, which required a generalization of the near-horizon asymptotic symmetry analysis to Einstein-Maxwell theory. We found that this yields an additional set of supertranslations that consists of an infinite-dimensional extension of gauge symmetries on which the horizon superrotations act in a non-trivial way. We discussed the physical interpretation of the symmetries and the associated Noether charges for electrically and magnetically charged Reissner-Nordstr\"{o}m black holes. Finally, we generalized the black hole memory effect to the Reissner-Nordstr\"{o}m black holes showing that a supertranslation at null infinity can be expressed as a composition of supertranslations, superrotations and large gauge transformations at the horizon.

Some questions remain open and require further study. The most important one is to find the correct interpretation of the horizon symmetries and their associated charges. While these quantities may be properly defined from the mathematical point of view, the physical interpretation of horizon symmetries and their local measures is not free of conceptual difficulties. On the one hand, in the quantum theory the horizon is a transitory state, which after having formed from gravitational collapse undergoes a Hawking process and eventually evaporates. On the other hand, the horizon is teleological. Therefore, any attempt to make sense out of it as a region of the spacetime that encodes relevant information of physical processes taking place in the bulk might be puzzling. Despite these circumstances, the explicit evaluation of the horizon charges seem to capture relevant information \cite{Donnay:2015abr, Hawking:2016msc, Donnay:2016ejv, Hawking:2016sgy} and it is worthwhile to keep investigating its mathematical and physical properties.

\subsection*{Acknowledgments}

LD and AP acknowledge support from the Black Hole Initiative at Harvard University, which is funded by a grant from the John Templeton Foundation. LD was also supported by a Fellowship of the Belgian American Educational Foundation and by the CNRS. L.D. and G.G. thank the Institute for Theoretical Physics of TU Wien and the Ecole Polytechnique for hospitality during their visits, where this work was partially done. The work of G.G. has been supported by NSF grant PHY-1214302. The work of H.G. is supported by the Austrian Science Fund (FWF), project P 28751-N2 and P 27182-N27.


\bibliographystyle{style}
\bibliography{references}

\end{document}